\documentclass[twocolumn,showpacs,%
  nofootinbib,aps,superscriptaddress,%
  eqsecnum,prd,notitlepage,showkeys,10pt]{revtex4-1}
  
\usepackage[utf8]{inputenc}
\usepackage{graphicx}
\usepackage{xcolor}
\usepackage{color,soul}
\usepackage{comment}
\usepackage{tabularx}
\usepackage{nameref}
\usepackage{varioref}
\usepackage{hyperref}
\usepackage{cleveref}
\usepackage{float}

\begin{document}
\title{RPS: Portfolio Asset Selection using Graph based Representation Learning}

\author{MohammadAmin Fazli}
\author{Parsa Alian}
\address{Computer Engineering, Sharif University of Tehran, Tehran, 1458889694, Iran
\\
\smallskip
Equally contributing authors.}
\email{Corresponding author: MohammadAmin Fazli. E-mail address: fazli@sharif.edu}
\thanks{No funding was received for conducting this study.}

\author{Ali Owfi}
\author{Erfan Loghmani}
\address{Computer Engineering, Sharif University of Tehran, Tehran, 1458889694, Iran}

\begin{abstract}
Portfolio optimization is one of the essential fields of focus in finance. There has been an increasing demand for novel computational methods in this area to compute portfolios with better returns and lower risks in recent years. We present a novel computational method called Representation Portfolio Selection (RPS) by redefining the distance matrix of financial assets using Representation Learning and Clustering algorithms for portfolio selection to increase diversification. RPS proposes a heuristic for getting closer to the optimal subset of assets. Using empirical results in this paper, we demonstrate that widely used portfolio optimization algorithms, such as MVO, CLA, and HRP, can benefit from our asset subset selection.
\end{abstract}

\maketitle

\section{Introduction}
Deciding how and where to invest money is one of the main challenges anyone with some savings faces. The complexity of finding an answer to this challenge has increased as the options for investment have grown in time, from gold and land to a myriad of financial assets like stocks, currencies, and cryptocurrencies. Moreover, investing money will be more important when the amount of money is considerable, such as for financial institutes, where any wrong decision or any fluctuation in the price of the invested asset can lead to considerable losses. In this situation, each individual can decide how to invest their money based on different parameters like knowledge, beliefs, and future predictions.

Two crucial concepts help investors diversify their assets. First is constructing an initial portfolio, which is a collection of financial assets that one individual holds. The second is managing the portfolio afterward. While these two concepts are interlinked, the dynamics of assets and trading constraints in markets introduce more complexities than the former, making them two different problems to solve. This paper will focus on the first problem, portfolio construction, and leave analyzing the complexities of portfolio management to future studies. Studies on the portfolio optimization problem have a long history. One of the first and most famous theories that studied portfolio construction was Markowitz's Portfolio Theory \cite{markowitz52, markowitz59, markowitz87}, which became the cornerstone of modern portfolio theory. In his study, he utilized the covariance matrix of the financial assets to define the problem as a quadratic programming problem. The success of Markowitz's theory embarked on creating many other methods in this field with different ideas. However, most portfolio construction and optimization methods have used the core idea of Markowitz's theory: the lower the correlation of a portfolio's assets, the lower the risk. Although naturally, many papers started studying the shortcomings of Markowitz's methods and addressed solutions to those shortcomings \cite{eltongruber95, king93, konnoyamazaki91, milis97, mitra2003review, rockafellar2000optimization}.

One of the Markowitz issues that we mitigate in this paper is the cardinality constraint problem. Markowitz’s method outputs the fraction of money to invest on each asset, with the assets and price history as the input. There are thousands of various assets available for investment, which can lead to hardships since managing such portfolios can be complicated. Moreover, it can be shown that limiting assets to a maximum count is a form of the Knapsack problem, which is computationally NP-hard \cite{garey-johnson79}. Different methods use different types of heuristics to solve this issue. For example, Crama and Schyns \cite{crama-schyns03} and Maringer \cite{maringer05} used simulated annealing, while others use clustering methods like k-means and hierarchical clustering \cite{lemieux2014clustering, raffinot2017hierarchical, leon2017clustering}. A whole family of other solutions formed based on Mantegna’s Minimum Spanning Tree method \cite{mantegna99book, mantegna99paper}. Some other methods try to reformulate the problem and solve it mathematically, such as Cesarone et al. \cite{Cesarone2013ANM}.

With the growing influence of computational and machine learning methods on financial markets, portfolio construction can be seen as an intermediary field from another perspective. A variety of computational methods have started to rise. Some methods first predict each asset changes in the future and use the results for portfolio construction \cite{ta2018prediction, chen2021mean, ma2021portfolio}. Other methods have also been used to directly construct the portfolios by training on previous data, such as Reinforcement Learning \cite{yu2019model}, and Deep Learning \cite{zhang2020deep}. A core characteristic of the asset markets, in general, is their complexity. Having a robust way to deal with this complexity will be useful to solve the portfolio optimization problem. One of the machine learning-based methods that can be used to do so is Representation learning, which is a learning method that embeds the entities in the problem in a low-dimensional feature space. For instance, \cite{du2020stock} uses text data like news articles related to each stock to design an embedding for stocks, then using clustering in the embedding space, the method can identify groups of stocks that had similar price behavior as well as similar news. On the other hand, \cite{hu2018deep} uses candlestick images and image neural network architectures to obtain a proper representation for each stock.

A new family of representation learning methods has formed in recent years, which embed the nodes of the graphs using their connections. This paper addresses the portfolio construction problem using the Node2Vec \cite{grover2016node2vec} method, a graph representation learning algorithm, on asset similarity graphs using the correlation matrix of assets. We leverage these representations in a two-phased portfolio optimization setting. First, we select a subset of assets and then weigh the obtained assets. While many portfolio optimization methods do not select the assets explicitly before weighing them, we focus on portfolio selection in this paper. We will indicate that better portfolios with higher returns and less risk can be achieved by separating the portfolio asset selection phase from the portfolio weighting phase. Furthermore, we state that doing so would help us to overcome multiple issues. Firstly, it eliminates the covariance estimation inaccuracy which exists in previous portfolio optimization methods. Secondly, it resolves portfolio optimization algorithms such as Mean-Variance Optimization (MVO) \cite{erlich2010MVO}, which would have convergence problems if given an extensive benchmark of assets by pre-selecting a heuristically optimal subset of assets.

 The paper is organized in the following structure.  Section II provides an in-depth description of the proposed method.  Section III discusses the other baseline methods that can be used for the portfolio selection problem.  At last, the results of the evaluation on multiple datasets are provided in section IV.

\subsection{Related works}

After the initial widespread success of Markowitz’s technique, lots of other approaches emerged in this area to enhance portfolio construction by modifying elements of the Markowitz technique or utilizing different strategies. Some attempts tried to improve the covariance estimation mechanism \cite{ledoit2004well, wong2003efficient, ledoit2004well}. Other methods have considered higher-order moments to capture the relationships between assets better \cite{maringer2009global, khan2020sustainable}.

Instead of improving portfolio optimization methods directly, some methods approached the problem by first selecting a subset of assets and then weighting them to construct the portfolio. Mantegna \cite{mantegna99paper} was one of the works that did so by using graphs to suggest a subset of assets. By finding an MST of the weighted graph, the method could find the hierarchical structure between stocks. Moreover, other combinatorial optimization methods on graphs are used to identify portfolios, as in Boginski et al. \cite{boginski2014network}, in which finding relaxed cliques in the network helps identify similar stocks. Clustering techniques have also found their way into financial studies. In De Prado \cite{de2016building}, authors use a hierarchical clustering approach to cluster stocks based on distances of their corresponding rows in the covariance matrix. Also, Kumari et al. \cite{kumari2019mean} uses a k-means clustering approach to identify stock groups with the same characteristics.

Recently, machine learning methods have found their ways in financial market studies \cite{henrique2019literature}. Portfolio optimization was not an exception for this trend. Some methods in this area leverage optimization methods to overcome computational and estimation problems of portfolio optimization methods. \cite{still2010regularizing} uses a regularization method to construct weights that are stable and robust to fluctuations. \cite{perrin2020machine} investigates optimization methods to find a method that could be practically used for real-world problems with lots of assets. Some other studies use prediction models first to predict each asset's future dynamics and then use the predictions to find optimal assets. \cite{ta2018prediction, chen2021mean, ma2021portfolio} investigate a range of machine learning methods from simpler methods (Linear Regression \& Support Vector Regression) to more complex methods (XGBoost \& LSTMs \footnote{Long Short-Term Memory}). Reinforcement Learning (RL) methods have also been studied, in which an agent learns how to construct appropriate portfolios by behaving in the environment \cite{yu2019model}.

Due to the recent advances in deep learning methods, methods that try to use deep embeddings for stocks are also developed. For instance, Du \& Tanaka-Ishii \cite{du2020stock} uses text data like news articles related to each stock to design an embedding for stocks, then using clustering in the embedding space, the method can identify groups of stocks that had similar price behavior as well as similar news. On the other hand, Hu et al. \cite{hu2018deep} uses candlestick images and image neural network architectures to reach a proper representation for each stock.

While these representation learning methods use text and image data, to the best of our knowledge, no previous method uses the graph of relationships between stocks to derive embeddings for stocks.

\section{Method}

This paper aims to present a novel utilization of representation learning as a diversification heuristic for a portfolio. The main idea is to choose uncorrelated assets for a portfolio to minimize the portfolio variance, which means less risk for the portfolio.
To elaborate, take $N$ assets in a portfolio which are pairwise uncorrelated, i.e., $\forall i \neq j, \rho_{i j}=0 .$ Then the variance for this portfolio would be 

$$
Var\left(P\right)=\sum_{i=1}^{N}\left(w_{i}\right)^{2} \sigma_{i}^{2}
$$

And to better show the idea, with the simplifying assumption of equal weights for the assets, i.e., $i, w_{i}=1 / N,$ we will have

$$
Var\left(P\right)=\frac{1}{N^{2}} \sum_{i=1}^{N} \sigma_{i}^{2} \leq \frac{\sigma_{M}^{2}}{N}
$$
where $$ \sigma_{M}=\max \left[\sigma_{i}: i=1, \ldots, N\right]$$

And as $N$ grows to infinity, $Var\left(P\right)$ will lean towards $0$. This shows us that the more pairwise uncorrelated assets we have in our portfolio, the smaller the portfolio's risk will be. Even though we do not have completely uncorrelated assets in real life, we can still build upon the shown idea and choose rather uncorrelated assets to reach a well-diversified low-risk portfolio. In this paper, we would want to build an augmented graph with some properties out of these assets based on their similarity or, to be more exact, their pairwise correlation, and then by applying a representation learning method on the graph, reach a new distance representation for the assets, which we can then choose the uncorrelated assets from to fulfill the idea shown above.

To use machine learning and deep learning methods on graph-structured data, we must first embed the graph or its nodes in a vector space. There are different types of methods being used for this purpose. Some of them use discrete metrics to define the similarity between nodes like the number of common neighbors \cite{ahmed2013distributed, cao2015grarep}. Others use random walks, and the probability of observing one node while starting from another node as the similarity of the two nodes \cite{perozzi2014deepwalk, grover2016node2vec}. These two methods are unsupervised and do not require any labels on data. However, if we have some labeled data, we can employ other methods that utilize the labels to find better representations. These methods have been improved by using other properties of graphs \cite{kipf2016semi, hamilton2017inductive}. Since our data is unlabeled and our initial metric (correlation) was not discrete, we chose to use Node2Vec and apply it on the correlation graph of the stocks that we constructed to reach a new similarity metric so that ultimately we can select a set of stocks which gives us a diversified portfolio.

To build the described graph, we first need to set weights for the edges. We chose Pearson correlation as the basis of the pairwise similarity measure for the assets, and then we had to adjust the measurement for better results. As for the adjustment and why it is needed since we will be applying Node2Vec to the graph and as it is a random walk based algorithm on the weights of our graph, at a given node, it will iterate through the edge $j $ with a probability of $ \frac{w_{j}}{\sum_{i=1}^{n} w_{i}}$. So we have to augment the basic Pearson Correlation such that the more correlated the two assets are, the less the weight of their intermediate edge will be and vice versa. After applying Node2Vec, starting a random walk on a node, results in more visits to uncorrelated assets. Furthermore, to ensure that we will not be visiting high correlated assets for a starting node in a Node2Vec with a walk length of greater than 1, we should also consider these properties in our transformation function:

\begin{itemize}
    \item 
    As the corresponding correlation of an edge approaches 0, the edge's weight should approach infinity.
    \item
    As the corresponding correlation of an edge approaches 1, the edge's weight should approach 0.
\end{itemize}

We used hyperbolic cotangent for our weight adjustment as it supported such properties. Then, The new redefinition of pairwise similarity between assets which was also used as the weight of the edges in our graph is calculated this way:

\begin{equation}
w_{ij} = \mid \coth(Correlation(i,j)) \mid -\coth(1)
\end{equation}

\begin{figure}[]
    \centering
    \includegraphics[width=0.5\textwidth]{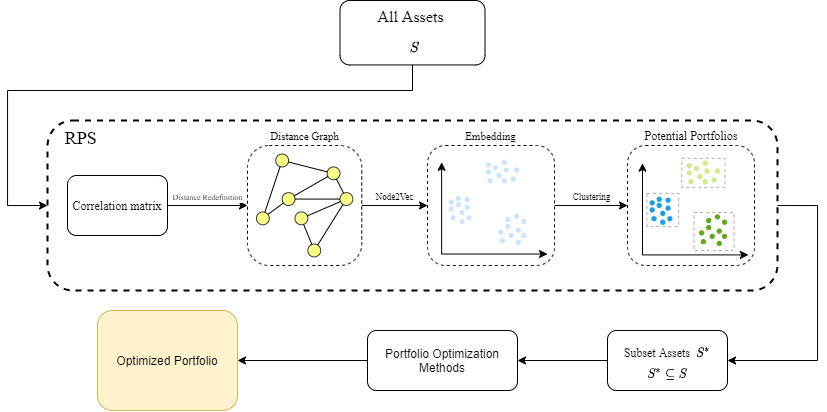}
    \caption{Stages of RPS algorithm.}
    \label{fig:methoddiagram}
\end{figure}

As our graph is built, the next step is to run Node2Vec on our graph. For each node, multiple random walks are executed based on the weights of its edges. That is, in each step, the algorithm visits the neighbor $j$ of a starting node with a probability of $ \frac{w_{j}}{\sum_{i=1}^{n} w_{i}}$, where $w_j$ is the weight of the edge between the starting node and neighbor $j$, and the algorithm iterates over the graph in this manner $l$ times for each execution, where $l$ is the walk-length of the algorithm. In the end, a new pair-wise distance representation is created based on the nodes visited during the execution of the algorithm for each starting node. Since we built the graph so that the edges with corresponding uncorrelated nodes to have greater weight than correlated ones, we will visit uncorrelated nodes for each node after this algorithm.

After embedding the nodes of the graph, we are free to use different clustering algorithms to reach our final portfolios. We used two different clustering methods. The first one was the k-means clustering algorithm, and the second one was the fuzzy c-means algorithm, which allows each node to be present in more than one subset. Our initial belief was that this relaxation might allow the procedure to achieve better performance. Since our representation of the graph is produced so that the uncorrelated assets are closer to each other, we will reach clusters of relatively uncorrelated assets, which can be interpreted as selected assets for our diversified portfolio.

The last step in the method is to input the resulted portfolios into a portfolio optimization method to calculate the fraction of wealth to invest in each of them. In this paper, we made use of three different optimization methods: MVO \cite{erlich2010MVO}, Hierarchical Risk Parity (HRP) \cite{pfitzinger2019HRP}, and Critical Line Algorithm (CLA) \cite{niedermayer2010CLA}.

\section{Evaluation Benchmark}

\paragraph{Minimum Spanning Tree (MST) and Hierarchical Clustering} The approach described in \cite{mantegna99paper} uses Kruskal's algorithm to build an MST over the complete graph of the market. The edges' weights are determined by the relation below.

\begin{equation}
d_{ij} = \sqrt{2(1 - \rho_{ij})}
\end{equation} 

where $\rho_{ij}$ is the correlation between asset $i$ and asset $j$. This approach relaxes the market graph and makes it easier for further operations. The paper itself does not specify a way to reach subsets of assets. We use Louvain Clustering Algorithm \cite{Blondel_2008} to extract smaller subsets of the market for evaluation. The clusters can be given to a weighting method, similar to RPS.

\paragraph{Graph Splex} The Graph Splex \cite{splex} tries to reach a diversified portfolio by creating a clique-like substructure of the market assets. We implemented this method using the pseudo-code described in the body of \cite{splex}. This subdivision can be weighted using the weight methods.

\paragraph{Simulated Annealing} Overall, various hill-climbing algorithms can be used as the solution for portfolio optimization cardinality problems, such as Simulated Annealing and Genetic Algorithms. The method starts with a random set of states for a subset of assets and changes the weights until a stable state is reached \cite{crama-schyns03, maringer05}.

\paragraph{Random Subsets} Since our goal is to constraint the cardinality of assets, we can use random divisions of the market as a baseline method. In this approach, Since random subdivision has access to every market subset, there is a probability that it can reach some of the best baskets. After selecting this subdivision, we can run the optimization methods on the portfolios.

\section{Results}

\subsection{Datasets}

To test our method in an empirical experiment, we chose three datasets from three different stock market indices in different timeframes.
Descriptions of the datasets that we used can be observed in Table \ref{tab:datasets}.
The S\&P 500 data is available in daily resolution
The Nikkei 225 and S\&P 100 datasets were obtained from Indextrack datasets \cite{BEASLEY2003621}. The prices in this dataset are indexed from 0 to 291, which are the weekly prices between March 1992 to September 1997, and the train and test ranges are expressed as an index value in Table \ref{tab:datasets}.

\begin{table}[]
    \centering
   \caption{Datasets Information}
    \begin{tabularx}{0.45\textwidth}{|*{4}{c}}
    \hline
    
    \multicolumn{1}{|p{1.86cm}|}{Index} & \multicolumn{1}{p{1.86cm}|}{Asset Count} & \multicolumn{1}{p{1.86cm}|}{Train Range} & \multicolumn{1}{p{1.86cm}|}{Test Range} \\ \hline
    
    \multicolumn{1}{|p{1.86cm}|}{S\&P 500} & \multicolumn{1}{p{1.86cm}|}{465} & \multicolumn{1}{p{1.86cm}|}{2019-04-01 to 2019-08-01} & \multicolumn{1}{p{1.86cm}|}{2019-08-02 to 2019-09-01} \\ \hline
    
    \multicolumn{1}{|p{1.86cm}|}{Nikkei 225} & \multicolumn{1}{p{1.86cm}|}{225} & \multicolumn{1}{p{1.86cm}|}{0 to 200} & \multicolumn{1}{p{1.86cm}|}{201 to 290} \\ \hline
    
    \multicolumn{1}{|p{1.86cm}|}{S\&P 100} & \multicolumn{1}{p{1.86cm}|}{98} & \multicolumn{1}{p{1.86cm}|}{0 to 200} & \multicolumn{1}{p{1.86cm}|}{201 to 290} \\ \hline
    
    \end{tabularx}
    \label{tab:datasets}
\end{table}

\subsection{Metrics}

We used two approaches to evaluate the performance of the algorithms used in our paper. We measured the future performances of their output portfolios using several financial metrics and assessed their stability both in time and against noise via computational methods.

The financial ratios which we used to evaluate the future performance of portfolios were as following:

\begin{itemize}

\item Correlation: Since our method's primary focus was to minimize the correlation between different assets' price values, we must evaluate how minimization of correlation in train data relates to the correlation of portfolios in the test data. All of the other measures are a byproduct of this value.

\item Return: The return of the portfolios was evaluated in the test range.

\item Risk: The risk of the portfolios is defined as the standard deviation of the asset returns in a given time range.

\item Sharpe Ratio:
\begin{equation}
    Sharpe\ Ratio = \frac{R_{p} - r_{f}}{\sigma_{p}}
\end{equation}

Where $R_{p}$ is the return of the portfolio, $r_{f}$ is the risk-free rate of return, and $\sigma{p}$ is the standard deviation of the portfolio.

\item
Information Ratio:
\begin{equation}
 Information\ Ratio = \frac{R_{p} - R_{b}}{\sigma_{R_{p} - r_{b}}}     
\end{equation}

Where $R_{p}$ is return of the portfolio, $R_{b}$ is return of the benchmark, and $\sigma_{R_{p} - r_{b}}$ is the standard deviation of the excess return.

\item
M2 Measure (Modigliani):
\begin{equation}
 M2\ Measure = SR * \sigma_{b} + r_{f}   
\end{equation}
Where $SR$ is the Sharpe Ratio, $r_{f}$ is the risk-free rate of return, and $\sigma_{b}$ is the standard deviation of the benchmark.

\end{itemize}

To evaluate the stability of the algorithms, we took another approach.

\newtheorem{dfn}{Definition}
\begin{dfn}
Suppose that the training phase of an algorithm has resulted in $k_1$ different portfolios, where each is a set of assets. If the model is trained again under different circumstances, the result of the train would be $k_2$ portfolios. A stability matrix ($SM$) is defined as a $k_1 \times k_2$ matrix where $SM_{ij}$ is the similarity value between portfolio $i$ of the first train phase and portfolio $j$ of the second train phase.
\end{dfn}

The similarity metric we used in this paper was Jaccard Similarity measures, which is defined as below between set $A$ and set $B$:

\begin{equation}
    JaccardSim(A, B) = \frac{|A \cap B|}{|A \cup B|}
\end{equation}

Where $|S|$ is the size of set $S$. We use a matrix instead of a list because we cannot determine an injective function between two phases of training, and therefore no direct mapping exists between two sets of portfolios. After forming the stability matrix, we can extract different measures from it. First of all, we calculate the maximum similarity value for each portfolio to find a mapping between phases. One problem in this process is that the count of portfolios can vary between phases. For example, since the Louvain clustering algorithm does not provide input for several clusters, the output cluster count might differ in different training sets. Furthermore, some of the weighting algorithms might not reach a conversion point for a specific portfolio, and therefore the subset would not be present in the training process results. As a result of this problem, the column-wise maximum of the matrix is not necessarily equal to the row-wise maximum. We combine the row-wise and the column-wise maximums before any further inspections to create a symmetric measure from the similarity matrix.

After taking the maximums, we use the average of maximums to compute the stability of the algorithm. As mentioned before, we also use two different stability tests:

\begin{itemize}
    \item Noise Stability: The stability of the method if a minuscule amount of Gaussian noise is applied to the correlation matrix of the assets.
    \item Time Stability: The method's stability if the time range of the training dataset is shifted for a small amount.
\end{itemize}

Note that these stability functions cannot be applied to Random and Simulated Annealing selection methods since they are statistical approaches for the optimization problem and unstable. No two consecutive runs with similar conditions would result in the same portfolio using these methods.

\subsection{Experiments}

\paragraph{Future Performance} To measure the performance of RPS for each of the datasets that we had, we constructed several portfolios using different methods and compared their performances via the metrics that we described above. Firstly, we built a set of portfolios with a two-phased approach that used RPS for their asset selection in their first phase, and then used one of the portfolio optimization methods CLA \cite{niedermayer2010CLA}, MVO \cite{erlich2010MVO}, or HRP \cite{pfitzinger2019HRP}. Then we created a set of other portfolios using our benchmark methods. Three out of four of these methods, Mantegna (MTN), Random (RND), and Graph Splex (SPX), can be used in a two-phased fashion like ours. After computing the subsets, we used the same portfolio optimization methods mentioned above. The Simulated Annealing (SA) method is a one-phased method, and the result of its training contains the weights and portfolios. Furthermore, RPS and MTN methods result in multiple portfolios. We ran RND and SA multiple times to compare all of the benchmarks, but the SPX approach outputs a consistent result for a given market. Lastly, we could not run the SPX method on Nikkei 225 and S\&P 500 datasets since it did not converge on our systems in a bounded time.

\begin{figure}[!ht]
    \label{fig:our_selection}
    \centering
    \includegraphics[width=0.46\textwidth]{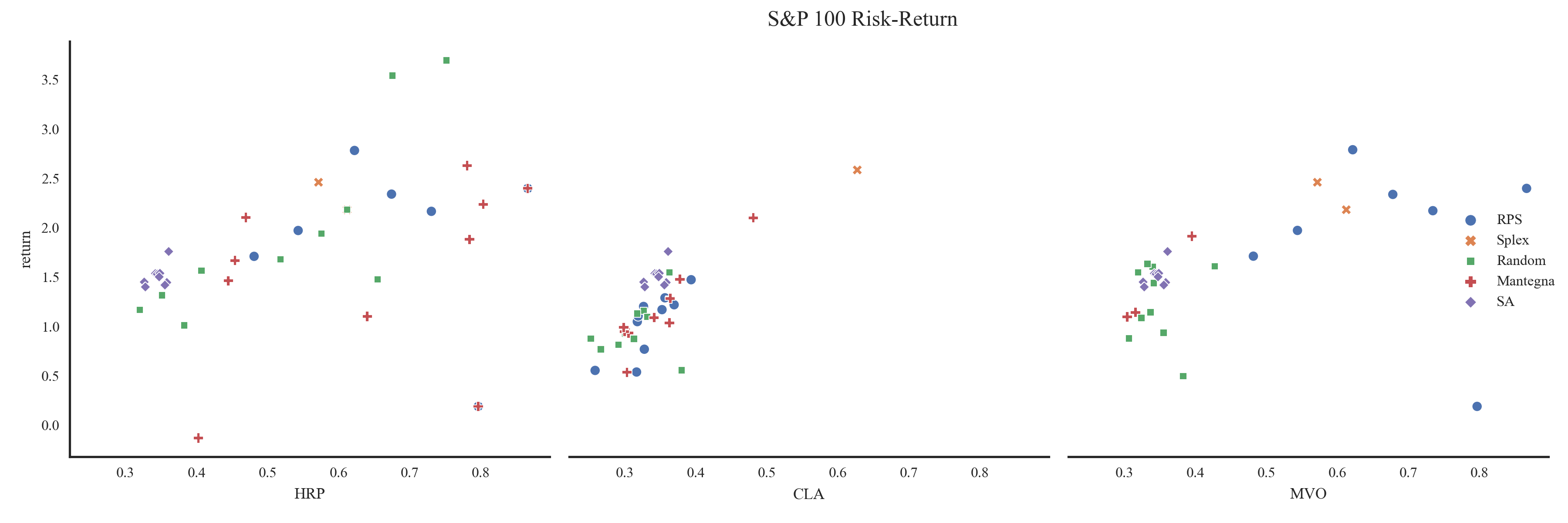}
    \includegraphics[width=0.46\textwidth]{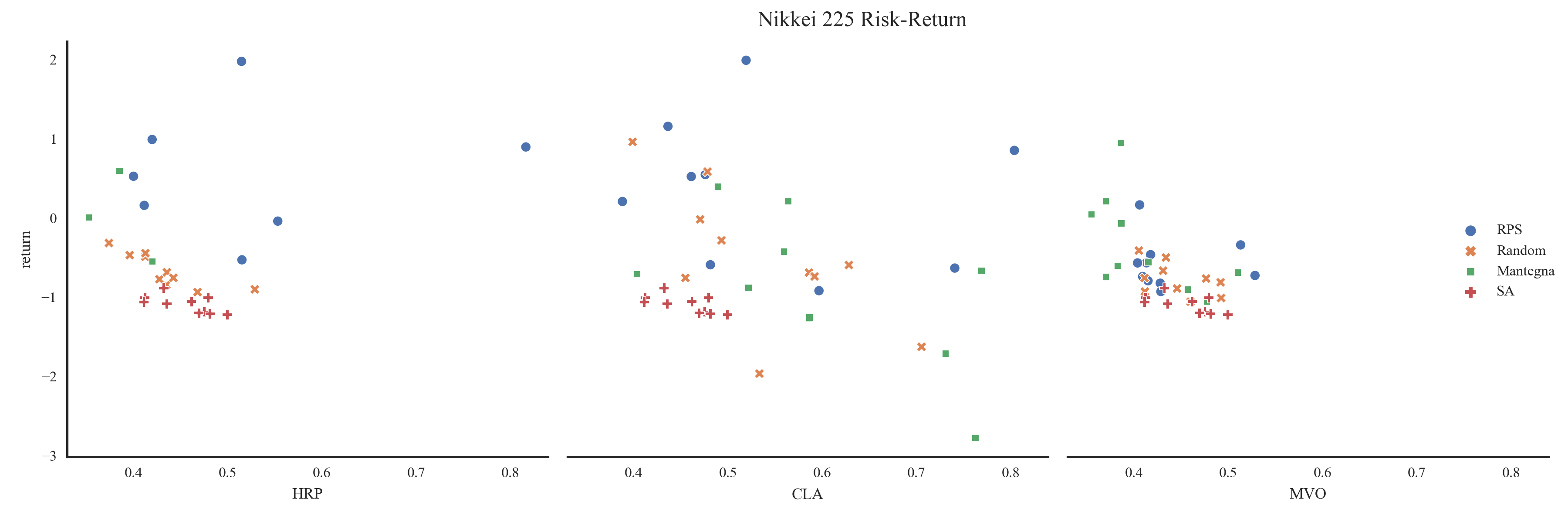}
    \includegraphics[width=0.46\textwidth]{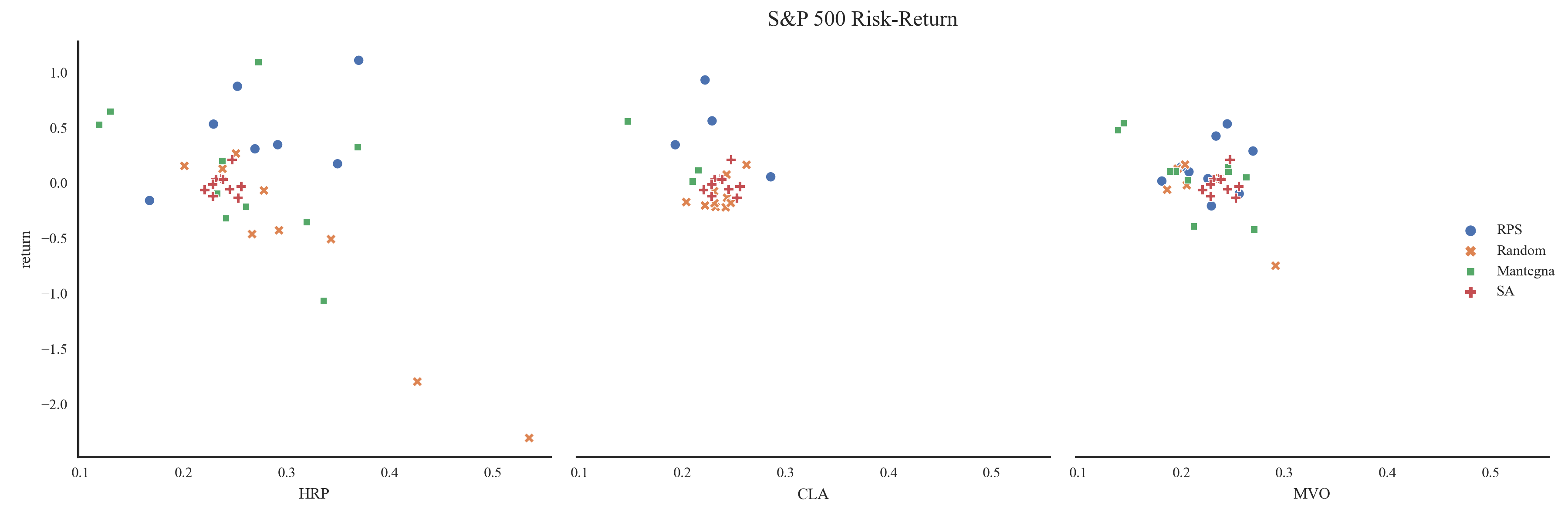}
    \caption{The efficient-frontier plot for the top 10 portfolios of the training set for RPS+optimization versus benchmark algorithms.}
\end{figure}

The top 10 portfolios in the training range of each method were then picked, and their performances were evaluated in the test range. Portfolios were sorted using Sharpe value to maximize the return while minimizing the risk. The risk-return graph depicting the discussed portfolios can be observed in Figure \ref{fig:our_selection}.

Moreover, the best value of future metrics for each of the methods are available in Tables \ref{tab:sp500fp}, \ref{tab:nikkeifp}, and \ref{tab:sp100fp}.

\begin{table*}[!t]
  \centering
  \caption{Future Performance Measures for S\&P 500 Dataset}
  \begin{tabular}{ccccccc}
    \hline
    Method & Correlation & Return & Risk & Sharpe Ratio & Information Ratio & M2 \\
    \hline
    Vanilla CLA & 0.162075 & 0.027747 & 0.114838 & 0.164118 & 0.242098 & 0.027636  \\
    RPS+CLA & 0.138791 & \textbf{1.110586} & 0.369897 & 3.435445 & \textbf{4.721627} & 0.817560 \\
    MTN+CLA & 0.196280 & 1.094718 & 0.369051 & \textbf{4.948828} & 4.654215 & \textbf{1.173791} \\
    RND+CLA & 0.171342 & 0.267575 & 0.535209 & 1.031657 & 1.140248 & 0.251739 \\
    \hline
    Vanilla HRP & 0.162075 & 0.011727 & \textbf{0.095295} & 0.216459 & 0.103681 & 0.015811  \\
    RPS+HRP & \textbf{0.008178} & 0.535214 & 0.269456 & 2.153089 & 2.277263 & 0.515710 \\
    MTN+HRP & 0.196280 & 0.541221 & 0.270475 & 3.698830 & 2.302783 & 0.879557  \\
    RND+HRP & 0.191148 & 0.166312 & 0.291356 & 0.773329 & 0.710050 & 0.190932 \\
    \hline
    Vanilla MVO & 0.162075 & 0.027747 & 0.114838 & 0.164118 & 0.242098 & 0.027636  \\
    RPS+MVO & 0.138791 & 0.932556 & 0.285581 & 4.163090 & 3.965298 & 0.988838 \\
    MTN+MVO & 0.235134 & 0.557674 & 0.215301 & 3.736161 & 2.372682 & 0.888345 \\
    RND+MVO & 0.174462 & 0.165436 & 0.262077 & 0.597288 & 0.706327 & 0.149494 \\
    \hline
    SA & 0.167465 & 0.273648 & 0.210593 & 1.117717 & 1.166047 & 0.271996  \\ \hline
  \end{tabular}
  \label{tab:sp500fp}
\end{table*}

\begin{table*}[!t]
  \centering
  \caption{Future Performance Measures for Nikkei 225 Dataset}
  \begin{tabular}{ccccccc}
    \hline
    Method & Correlation & Return & Risk & Sharpe Ratio & Information Ratio & M2 \\
    \hline
    Vanilla CLA & 0.450845 & 0.143052 & 0.436134 & 0.261482 & 0.278425 & 0.142717  \\
    RPS+CLA & \textbf{0.094671} & \textbf{1.991665} & 0.387936 & 3.818792 & \textbf{4.599676} & 1.665158 \\
    MTN+CLA & 0.210820 & 0.395800 & 0.403417 & 0.790543 & 0.920129 & 0.351768 \\
    RND+CLA & 0.189425 & 0.962127 & 0.398781 & 2.390352 & 2.225895 & 1.045625 \\
    \hline
    Vanilla HRP & 0.450845 & 0.042383 & 0.422741 & 0.069599 & 0.081713 & 0.044518  \\
    RPS+HRP & 0.176072 & 0.165529 & 0.403649 & 0.385897 & 0.389200 & 0.176268 \\
    MTN+HRP & 0.210820 & 0.946481 & 0.354701 & 2.427145 & 2.189820 & 1.061583 \\
    RND+HRP & 0.265348 & -0.411154 & 0.405019 & -1.037120 & -0.940445 & -0.440912 \\
    \hline
    Vanilla MVO & 0.450845 & 0.143052 & 0.436134 & 0.261482 & 0.278425 & 0.142717  \\
    RPS+MVO & \textbf{0.094671} & 1.978367 & 0.400116 & \textbf{3.826142} & 4.569014 & \textbf{1.668345} \\
    MTN+MVO & 0.210820 & 0.594692 & \textbf{0.352680} & 1.520505 & 1.378710 & 0.668362 \\
    RND+MVO & 0.105741 & -0.315837 & 0.374047 & -0.868171 & -0.720674 & -0.367636 \\
    \hline
    SA & 0.558230 & -0.937416 & 0.420207 & -2.070724 & -2.153834 & -0.889199 \\
    \hline
  \end{tabular}
  \label{tab:nikkeifp}
\end{table*}

\begin{table*}[!t]
  \centering
  \caption{Future Performance Measures for S\&P 100 Dataset}
  \begin{tabular}{ccccccc}
    \hline
    Method & Correlation & Return & Risk & Sharpe Ratio & Information Ratio & M2 \\
    \hline
    Vanilla CLA & 0.418337 & 1.269428 & 0.198022 & 4.348930 & 4.385291 & 1.262862 \\
    RPS+CLA & 0.177503 & 2.783124 & 0.481028 & 4.457300 & 9.635014 & 1.294110 \\
    MTN+CLA & 0.155783 & 2.630134 & 0.402634 & 4.461950 & 9.104422 & 1.295450 \\
    SPX+CLA & 0.870104 & 2.461136 & 0.571726 & 4.289180 & 8.518312 & 1.245634 \\
    RND+CLA & 0.208234 & \textbf{3.696725} & 0.320223 & \textbf{5.226347} & \textbf{12.803521} & \textbf{1.515855} \\
    \hline
    Vanilla HRP & 0.418337 & 1.278512 & 0.195795 & 4.449693 & 4.416796 & 1.291916 \\
    RPS+HRP & 0.255930 & 1.473516 & 0.257972 & 3.724968 & 5.093098 & 1.082950 \\
    MTN+HRP & 0.155783 & 2.100452 & 0.298370 & 4.351074 & 7.267409 & 1.263480 \\
    SPX+HRP & 0.870104 & 2.585765 & 0.627108 & 4.109122 & 8.950544 & 1.193717 \\
    RND+HRP & 0.206733 & 1.548159 & \textbf{0.252117} & 4.243590 & 5.351972 & 1.232489 \\
    \hline
    Vanilla MVO & 0.418337 & 1.269428 & 0.198022 & 4.348930 & 4.385291 & 1.262862 \\
    RPS+MVO & 0.177503 & 2.790539 & 0.481578 & 4.476027 & 9.660732 & 1.299509 \\
    MTN+MVO & 0.339761 & 1.914374 & 0.303960 & 4.823667 & 6.622061 & 1.399747 \\
    SPX+MVO & 0.870104 & 2.461136 & 0.571726 & 4.289180 & 8.518312 & 1.245634 \\
    RND+MVO & \textbf{0.043606} & 1.634323 & 0.306306 & 4.889193 & 5.650800 & 1.418641 \\
    \hline
    SA & 0.217874 & 1.759462 & 0.326334 & 4.851917 & 6.084802 & 1.407893 \\
    \hline
  \end{tabular}
  \label{tab:sp100fp}
\end{table*}

\begin{figure}[!ht]
    \centering
    \includegraphics[width=0.47\textwidth]{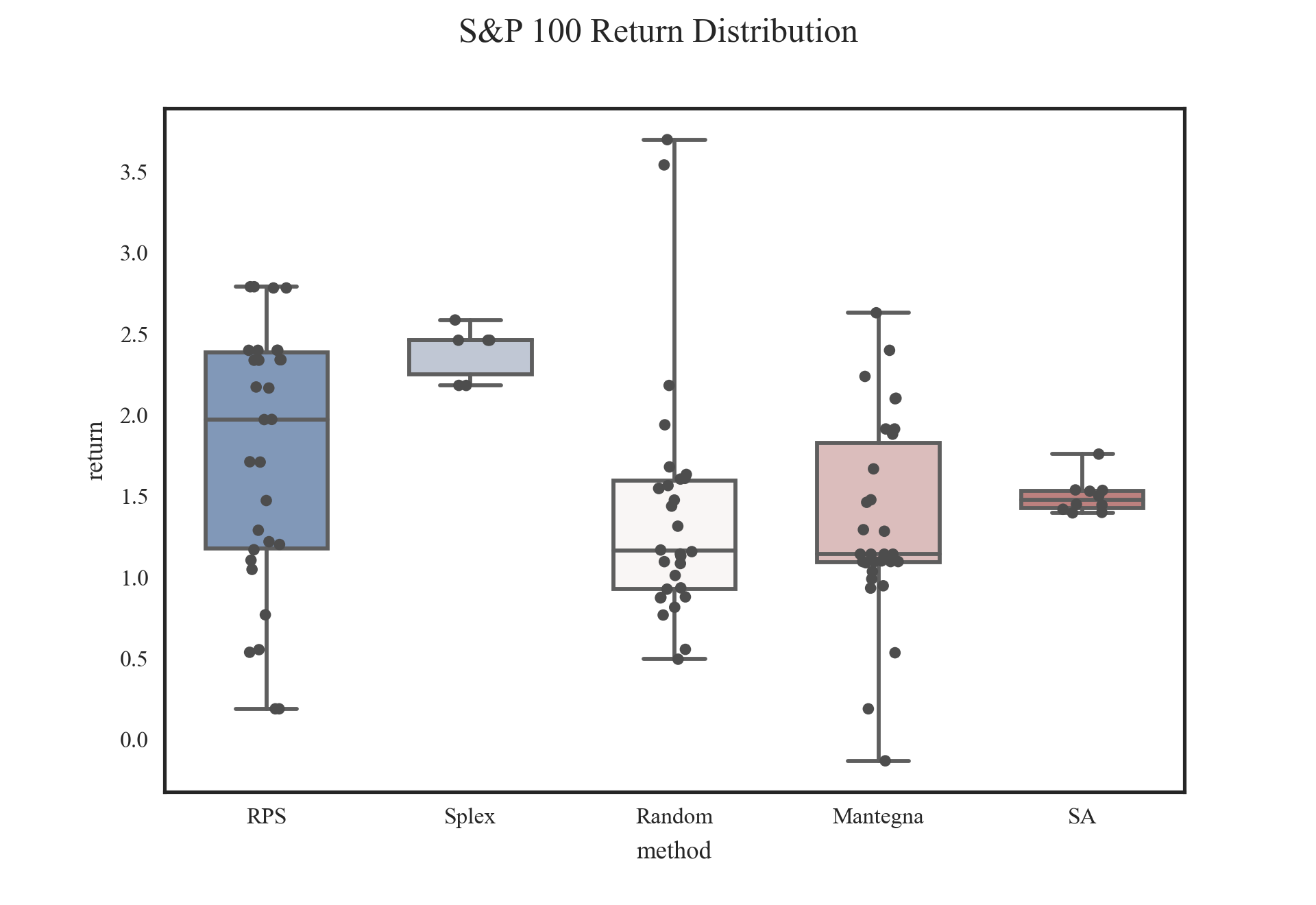}
    \includegraphics[width=0.47\textwidth]{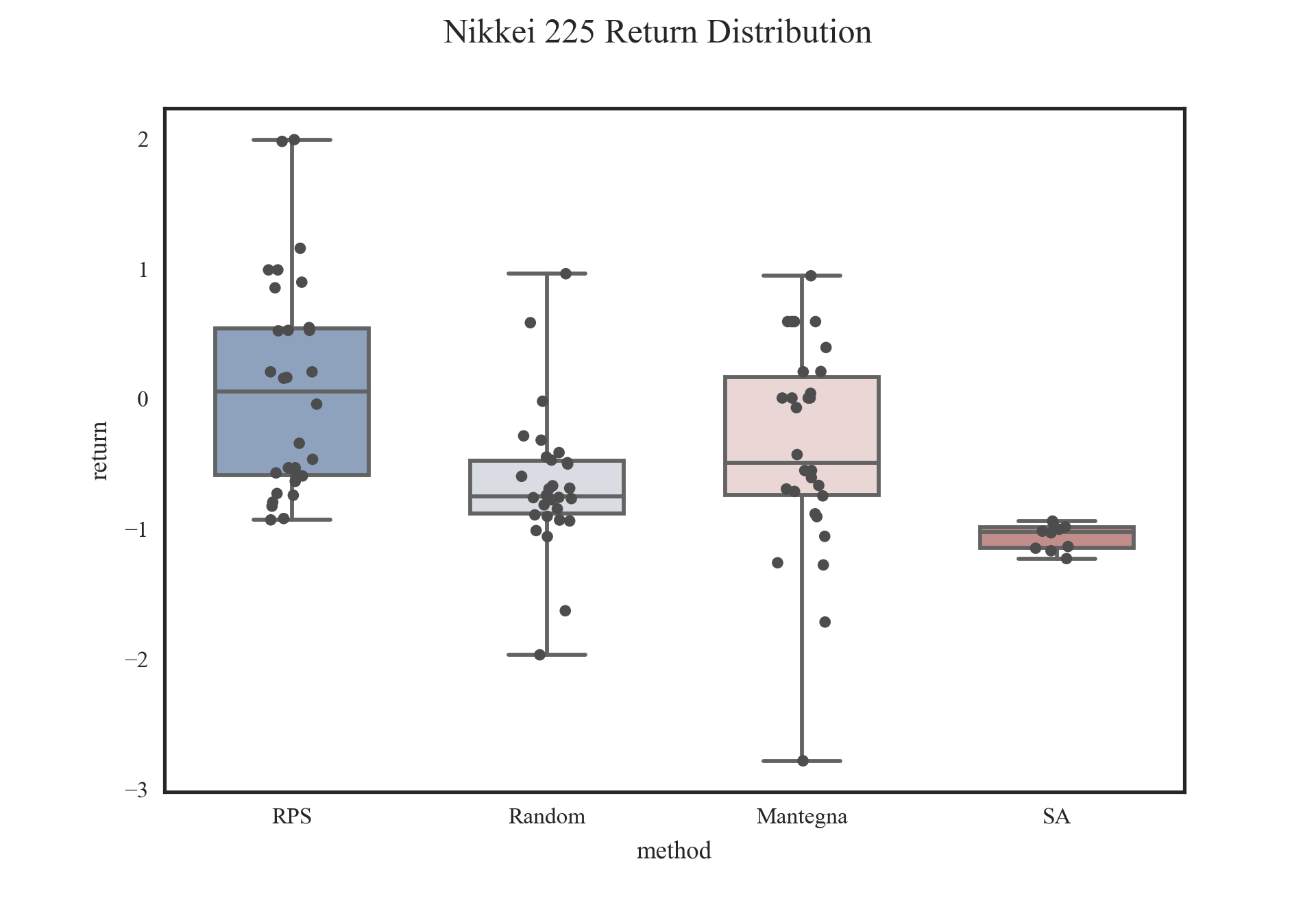}
    \includegraphics[width=0.47\textwidth]{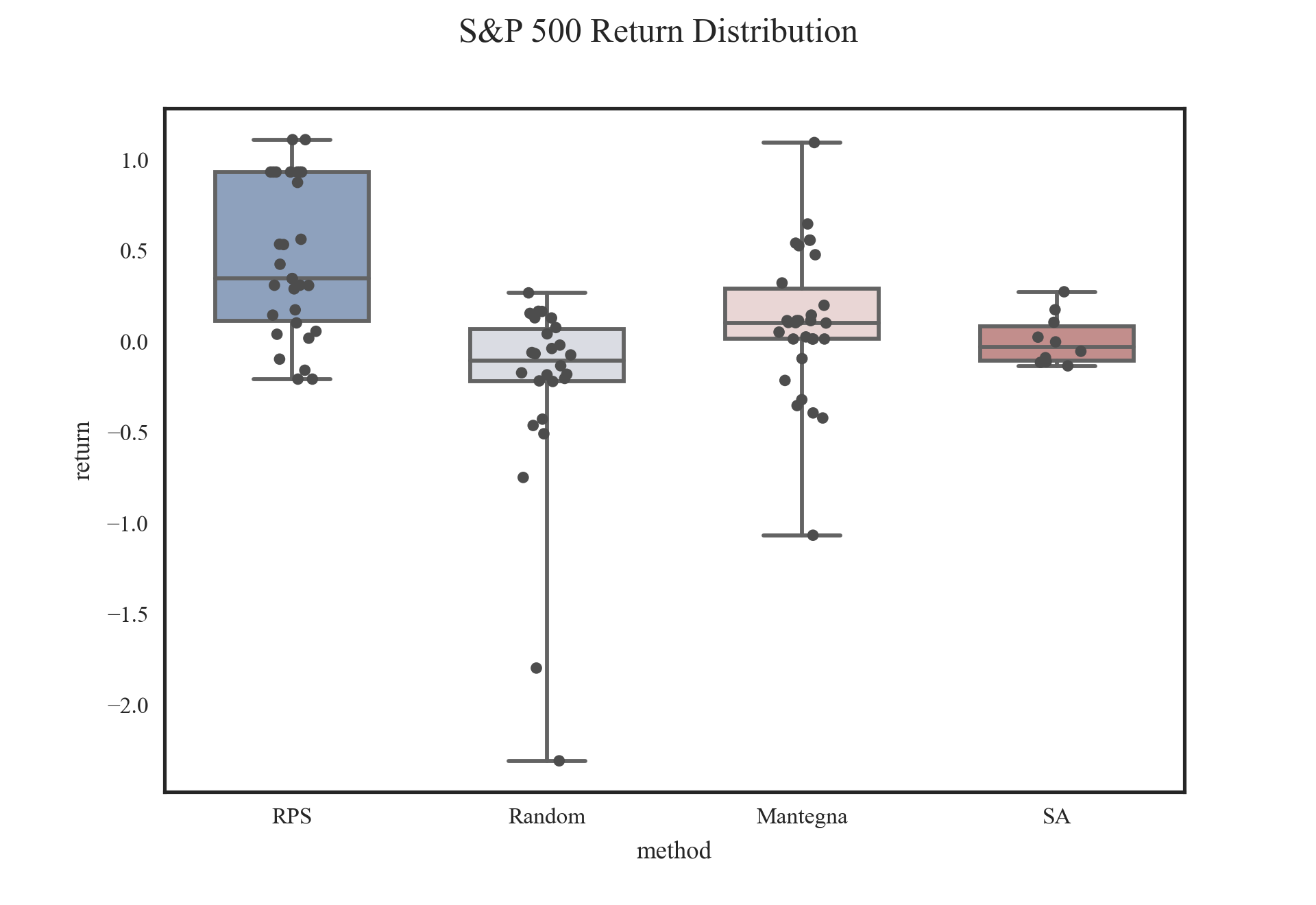}
    \caption{Return distribution}
    \label{fig:return_dist}
\end{figure}

It can be seen in these tables that in S\&P 500 and Nikkei 225, RPS has reached the best possible return among other methods and also the best correlation value. It can be seen that selection in all of the instances caused the correlation value to decrease from the original correlation of the assets. However, the results of performances are pretty different in S\&P 100 database. It is consistent with the notion that a random selection phase can lead to relatively high-performance portfolios since it can reach any point in the space of portfolios. The relative lousy performance of all other algorithms can be explained by looking at the correlation of all the assets in the dataset. The overall average correlation between assets of S\&P 100 is equal to 0.418337. Since the market assets are highly correlated, it becomes harder for the selection algorithms to find a highly uncorrelated subset, and therefore, most of the portfolios would move in the direction of the general trend. Both good performance of selection algorithms and bad performance of random selection in Nikkei 255 and bad performance of selection algorithms and good performance of random selection in S\&P 100 can be explained using this argument. Moreover, we can inspect the return of all portfolios in the markets in Figure \ref{fig:return_dist}.

It can be seen in Figure \ref{fig:return_dist} that RPS achieves a higher average return. In S\&P 100, few of the RND portfolios were able to reach a high return value, and also SPX achieves a higher average return, but as can be seen in the figure, RPS can obtain higher average performance.

\paragraph{Stability} The time and noise stability metrics are shown in table \ref{tab:ns} and \ref{tab:ts}. Gaussian noise with $\mu=0$ and $\sigma=0.01$ was applied to the market correlation matrix for noise stability. For time stability, the train time ranges were shifted to 20 data points for each dataset.

\begin{table}[!hbtp]
  \centering
  \caption{Noise Stability}
  \begin{tabular}{cccc}
    \hline
    Method & S\&P 100 & Nikkei 225 & S\&P 500 \\ \hline
    Mantegna & \textbf{0.528711} & \textbf{0.479277} & \textbf{0.427159} \\
    RPS & 0.460720 & 0.373794 & 0.085135 \\
    Splex & 0.000000 & - & - \\
    \hline
  \end{tabular}
  \label{tab:ns}
\end{table}

\begin{table}[!hbtp]
  \centering
  \caption{Time Stability}
  \begin{tabular}{cccc}
    \hline
    Method & S\&P 100 & Nikkei 225 & S\&P 500 \\ \hline
    Mantegna & 0.336598 & \textbf{0.360965} & \textbf{0.228190} \\
    RPS & \textbf{0.557390} & 0.231124 & 0.119387 \\
    Splex & 0.400000 & - & - \\
    \hline
  \end{tabular}
  \label{tab:ts}
\end{table}

In noise stability metrics, MTN was able to outperform RPS and SPX methods. However, the margin of superiority was not significant in the S\&P 100 dataset, resulting from the smaller size of the database and the lower number of options available in the selection phase. As the size of the dataset grows, stability drops for all of the methods.

\begin{figure}[!ht]
    \centering
    \includegraphics[width=0.47\textwidth]{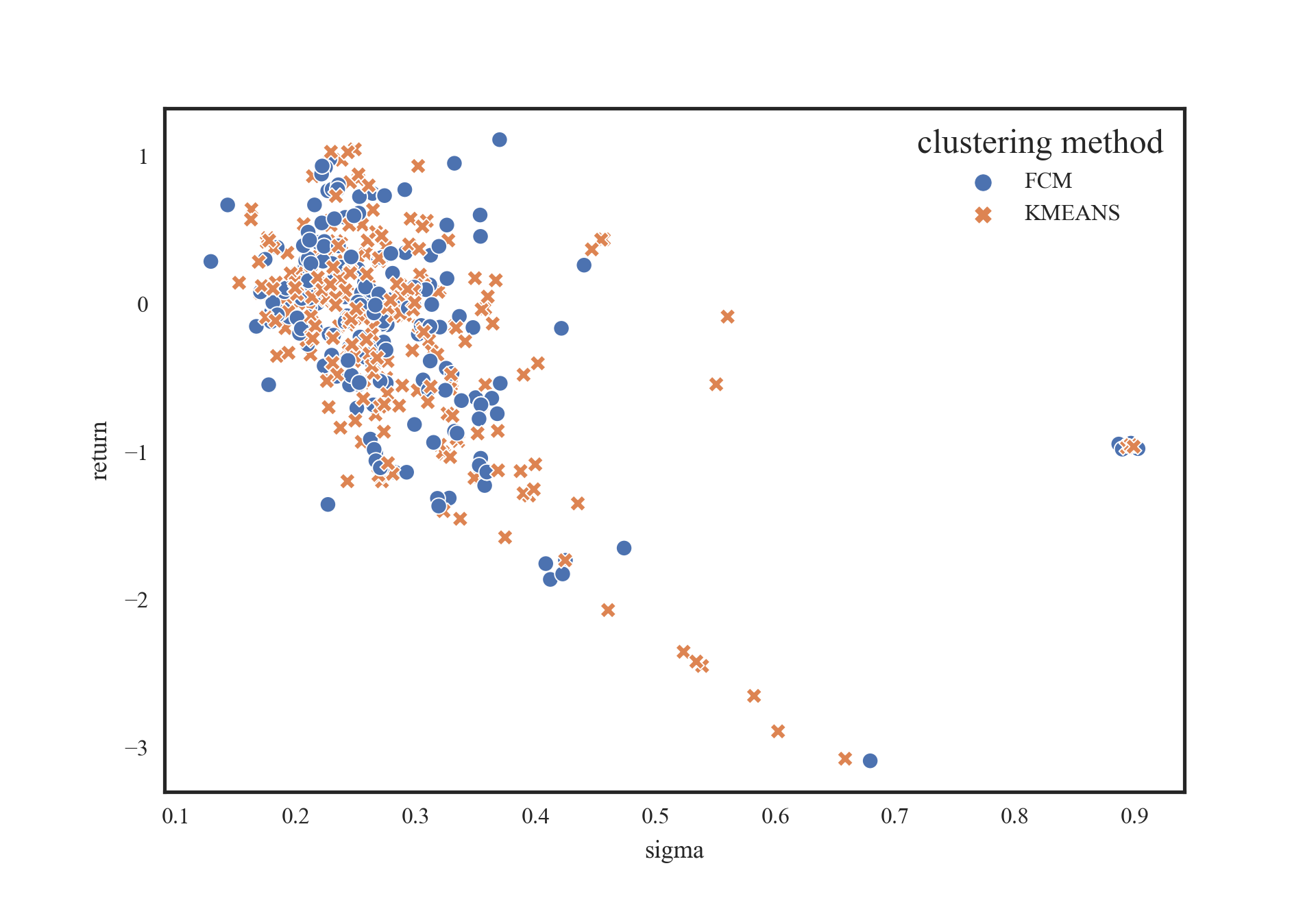}
    \caption{Risk-Return distribution of RPS on S\&P 500 dataset, separated by clustering method.}
    \label{fig:rps_cluster}
\end{figure}

\paragraph{RPS Clustering Method} Another issue to analyze is whether the clustering method (k-means or fuzzy c-means) affects outcoming portfolios or not. Figure \ref{fig:rps_cluster} depicts the risk-return relation for all different setups of RPS ran in the test range.

As seen in this figure, there are no significant differences between the portfolios using these two clustering methods.

\section{Conclusion}

This paper aimed to provide a new portfolio selection method that could outperform other algorithms used in the portfolio asset selection phase. To fulfill our goals, we provided a novel portfolio selection method based on representation learning and graph embedding called RPS. As seen in the results, various portfolio weight optimization methods can benefit from using RPS, since in every instance, using RPS selection improved the overall performance of the method versus the vanilla usage of that method. We have also seen that the portfolios constructed by RPS have a generally higher return than their rivals. Future directions can add dynamics to the problem and consider changing portfolios in time by periodic training RPS. Other weight optimization methods can also be tested on the selection to improve the portolios' performance further.

\section*{Code and Data Availability Statement} \label{s:data}
Implementation of all the prediction methods and the used data is available on
\url{https://github.com/parsaalian/RPS}

\section*{acknowledgements}
The authors did not receive support from any organization for the submitted work.

\section*{conflict of interest}
No funds, grants, or other support was received. All authors certify that they have no affiliations with or involvement in any organization or entity with any financial interest or non-financial interest in the subject matter or materials discussed in this manuscript.

\end{document}